\documentclass[conference]{IEEEtran}
\IEEEoverridecommandlockouts

\usepackage{cite}
\usepackage{amsmath,amssymb,amsfonts}
\usepackage{algorithmic}
\usepackage{graphicx}
\usepackage{textcomp}
\usepackage{xcolor}
\def\BibTeX{{\rm B\kern-.05em{\sc i\kern-.025em b}\kern-.08em
    T\kern-.1667em\lower.7ex\hbox{E}\kern-.125emX}}
\usepackage{multirow}

\begin{document}

\title{System Cards for \\ AI-Based Automated Decision Systems}

\author{
    \IEEEauthorblockN{Furkan Gursoy, Ioannis A. Kakadiaris}
    \IEEEauthorblockA{\textit{Computational Biomedicine Lab} \\
\textit{Dept. of Computer Science} \\
\textit{University of Houston}\\
Houston, TX, USA
    \\ \{fgursoy, ioannisk\}@uh.edu}
}

\maketitle
\thispagestyle{plain}
\pagestyle{plain}

\begin{abstract}
 Decisions impacting human lives are increasingly being made or assisted by automated decision-making algorithms. Many of these algorithms process personal data for predicting recidivism, credit risk analysis, identifying individuals using face recognition, and more. While potentially improving efficiency and effectiveness, such algorithms are not inherently free from bias, opaqueness, lack of explainability, maleficence, and the like. Given that the outcomes of these algorithms have a significant impact on individuals and society and are open to analysis and contestation after deployment, such issues must be accounted for before deployment. Formal audits are a way of ensuring algorithms meet the appropriate accountability standards. This work, based on an extensive analysis of the literature and an expert focus group study, proposes a unifying framework for a system accountability benchmark for formal audits of artificial intelligence-based decision-aiding systems. This work also proposes system cards to serve as scorecards presenting the outcomes of such audits. It consists of 56 criteria organized within a four-by-four matrix composed of rows focused on (i) data, (ii) model, (iii) code, (iv) system, and columns focused on (a) development, (b) assessment, (c) mitigation, and (d) assurance. The proposed system accountability benchmark reflects the state-of-the-art developments for accountable systems, serves as a checklist for algorithm audits, and paves the way for sequential work in future research.
\end{abstract}


\section{Introduction}

\begin{figure*}
    \centering
  \includegraphics[width=0.84\textwidth]{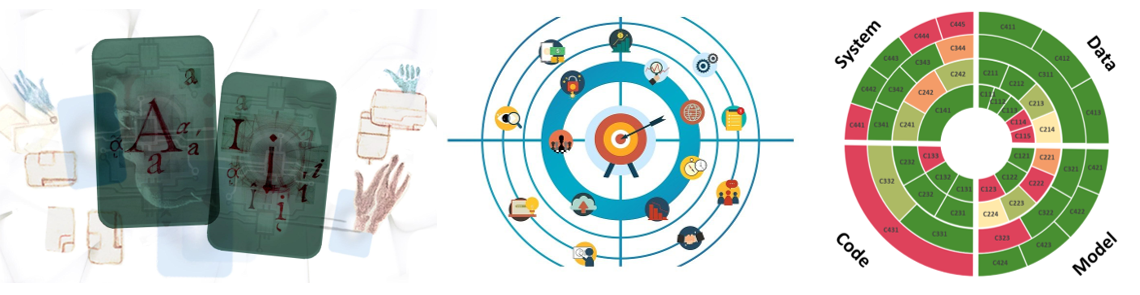}
  \caption{\textbf{Automated Decision Systems Meet System Accountability Benchmark to Generate System Cards.} From left to right, the three images represent an automated decision system, the system accountability benchmark, and the resulting system card, respectively. $f($Automated Decision Systems, System Accountability Benchmark$) = $ System Cards. (The leftmost image: Alina Constantin / Better Images of AI / Handmade A.I / CC-BY 4.0)}
  \label{fig:teaser}
\end{figure*}

The use of AI-based systems has expanded dramatically in recent decades. They currently play an active part in informing many critical decisions impacting human lives such as criminal justice decisions \cite{wexlerLifeLibertyTrade2018}, public education decisions \cite{AlgorithmicSystemsEducation}, and credit risk assessment \cite{bhatoreMachineLearningTechniques2020}. Algorithms, broadly defined as a process or set of rules to be followed in problem-solving operations, have long been used in point systems, formal decision models, and informal rules of thumb \cite{allisonBureaucraticPoliticsParadigm1972}. For example, in 1978, the District Attorney’s office in Harris County, TX developed a scoring system for determining whether a defendant should be held in jail while awaiting trial, using seven questions (e.g., previous offenses, family in area, and home phone ownership)\footnote{Based on an interview with a former employee who worked on the development of the scale (March 16, 2021).}. Defendants who scored higher on the scale were considered less risky to release while awaiting trial than those who scored lower.

While algorithms have long been used in decision-making, both the scope and complexity of such algorithms have been significantly increasing in recent years. The use of increasingly more complex algorithms that are usually based on AI raises legal challenges in balancing the protection of privacy, civil liberties, and intellectual property; social challenges in creating trust and maintaining accountability; design challenges in producing systems that are both efficacious and transparent; and policy challenges in incorporating algorithms into decision-making. The qualitatively and quantitatively different nature of the current wave of automated decision systems (ADSs) presents three challenges.

The first challenge is that with advances in data collection and machine learning, explicit algorithms have gained much broader acceptance than was previously possible. Most early research suggested that people were generally averse to computer algorithms, especially in human relations and moral decision-making \cite{dawesRobustBeautyImproper1979, dietvorstAlgorithmAversionPeople2015, gogollRageMachineAutomation2018, leeUnderstandingPerceptionAlgorithmic2018}. However, some research has suggested that this algorithm aversion has been dramatically declining \cite{kennedyTrustPublicPolicy2021, loggTheoryMachineWhen2017}. As trust in algorithms increases, so too does the danger of viewing decisions of these algorithms as "just math" \cite{RK_braynePredictSurveilData2020}, where even decisions that exacerbate racial bias and inequality are "technowashed" by their encoding in an algorithm \cite{RK_braynePredictSurveilData2020, winstonPioneerPredictivePolicing2018}.

The second challenge is that the new wave of ADSs is often impenetrable and, thus, less prone to evaluation. The bail hearing algorithm mentioned above, despite its flaws, is reasonably easy to understand and, therefore, to challenge legally. Today's algorithms are based on millions of data points, utilize hundreds of features, and model relationships using formulas that defy easy explanation. This makes them especially frustrating to challenge since, for instance, the exclusion of certain explicit factors like race can be accurately reconstructed from other factors in the model \cite{RK_oneilWeaponsMathDestruction2016}. Moreover, in part, because development teams for these systems tend to be quite homogeneous, the potential biases in such systems may not be readily apparent to team members \cite{broussardArtificialUnintelligenceHow2018}. In more complex applications, such as facial or voice recognition, challenging a software's assessment can be almost impossible. 

The third challenge is that AI-based systems often obscure the policy decisions and social factors underlying their design and application. The large datasets on which the machine learning models are based were formed within the context of historical and current social processes \cite{RK_benjaminRaceTechnologyAbolitionist}. Since machine learning solutions involve using historical data to create future forecasts, biases inherent in relevant social structures are reproduced in the algorithm \cite{RK_braynePredictSurveilData2020, lambrechtAlgorithmicBiasEmpirical2019}. It should not be surprising, for example, that predictive policing and sentencing algorithms are often accused of racial bias – since they are created using datasets produced by a racially biased criminal justice system, they reproduce those same patterns, even when the race is not explicitly used as a factor \cite{angwinMachineBias2016}. Algorithms can also produce devastating feedback loops. Pre-trial detention algorithms that (implicitly) incorporate information about a person's income can result in people already at the margins of the economy losing their employment and falling further into poverty \cite{RK_oneilWeaponsMathDestruction2016}. Predictive policing algorithms that send more officers to areas with a large number of low-level, non-violent arrests are likely to result in still more low-level, non-violent arrests \cite{RK_braynePredictSurveilData2020}. The process can quickly produce a negative feedback loop, punishing already disadvantaged members of society.

On the other hand, the use of AI-based systems in decision-making is not entirely negative. For example, current sentencing and pre-trial detention algorithms have often been adopted by states looking to decrease overall prison populations without being labeled “soft on crime.” Indeed, Kleinberg \textit{et al}. \cite{kleinbergHumanDecisionsMachine2018} demonstrate that following the advice of a defendant risk evaluation algorithm can reduce the number of people imprisoned by 42\% without increasing crime rates or reduce crime rates by 24.8\% without increasing prison populations. It must also be considered that the systems these algorithms are informing are traditionally not very accurate – many studies have found that professional judges are not particularly good at assessing defendant risk and are not very reliable in their judgments \cite{austinSurveyJudgesResponses1977, dhamiBailingJailingFast2001}.

As AI-based ADSs expand into even more areas of life, the discussion needs to move beyond the arguments about whether the use of such systems is good or bad and towards a discussion of how those systems can be made accountable in terms of our ability to evaluate, challenge, and even override their behavior. However, to the best of our knowledge, a framework that serves as an exhaustive system accountability benchmark covering the whole lifecycle and different aspects of the system and its components is lacking in the literature. This study addresses this challenging gap. Our contribution is the proposal of a novel system accountability benchmark and resulting system cards for machine learning-based automated decision systems (see Figure \ref{fig:teaser}).

The rest of the work is structured as follows. Section \ref{sec:bg} provides a review of selected literature and provides the background that drives the design and decisions for the proposed system accountability benchmark. Section \ref{sec:met} details the methodology used in developing and validating the proposed benchmark. Section \ref{sec:sc} presents the criteria within the system accountability benchmark and its organization. Section \ref{sec:dis} discusses several topics around accountability and the proposed benchmark.  Final remarks and directions for future research are provided in Section \ref{sec:co}.

\section{Background}\label{sec:bg}

There is an increasing number of studies that provide a framework or a discussion on algorithmic accountability including the relevant studies in algorithm auditing. Only a few of these studies have narrowed their focus to a single aspect or a dimension of algorithmic accountability. Many studies focus on the meaning of and the processes for algorithm accountability in general or attempt to address multiple dimensions and aspects within an accountability framework in a holistic manner. On the other hand, as the field is still in development, no individual study captures all aspects and dimensions that are worthy of discussion for such a framework.

A group of studies \cite{loiAccountabilityUseArtificial2021, cechAgencyForumMechanisms2021, wieringaWhatAccountWhen2020} largely concentrate on defining algorithmic accountability and what it means to be accountable for a decision-aiding system. 
Some studies \cite{henriksenSituatedAccountabilityEthical2021, slotaManyHandsMake2021} conduct interviews with relevant stakeholders to understand challenges around accountable artificial intelligence.
Another group of studies \cite{rajiClosingAIAccountability2020, mokanderEthicsBasedAuditingAutomated2021, zicariZInspectionProcessAssess2021} focus on elaborating the process of auditing such as how an algorithm should be audited in terms of associated tasks, actions, and steps. 
Few studies \cite{bandyProblematicMachineBehavior2021, wilsonBuildingAuditingFair2021} discuss actual applications of audits on algorithmic products. A relatively large number of studies \cite{ukAuditingMachineLearning, brownAlgorithmAuditScoring2021, koshiyamaAlgorithmAuditingSurvey2021, najaSemanticFrameworkSupport2021, tagiouToolSupportedFramework2019, krafftActionOrientedAIPolicy2021} are more extensive in scope and propose accountability frameworks that generally aim at capturing a diverse set of aspects and perspectives. 
The latter group also tends to present more tangible evaluation criteria rather than a relatively more abstract discussion of principles. On the other hand, some studies focus on particular themes such as nondiscrimination \cite{bartlettAlgorithmicAccountabilityLegal, wilsonBuildingAuditingFair2021, saleiroAequitasBiasFairness2019}, data \cite{bartlettAlgorithmicAccountabilityLegal, gebruDatasheetsDatasets2021}, models \cite{mitchellModelCardsModel2019}, transparency \cite{najaSemanticFrameworkSupport2021}, human rights \cite{mcgregorINTERNATIONALHUMANRIGHTS2019}, reputational concerns \cite{buhmannManagingAlgorithmicAccountability2020}, audits by everyday users \cite{shenEverydayAlgorithmAuditing2021}, and assurance \cite{batarsehSurveyArtificialIntelligence2021, kazimAIAssuranceProcesses2020}.
Finally, there are studies that examine regulations \cite{clarke2231116thCongress2019, GuidanceAIAuditing, EURLex52021PC0206EURLex, loiAccountabilityUseArtificial2021, clarkeAlgorithmicAccountabilityAct2022,
maccarthyExaminationAlgorithmicAccountability2019, kazimAIAuditingImpact2021, gursoyCriticalAssessmentAlgorithmic2022}.

An integrative literature review identified several concepts that are relevant for developing an accountability framework for AI-based ADSs. The criteria contained in the framework that is presented in the next section are driven by these broad concepts: sociotechnical context, interpretability and explainability, transparency, nondiscrimination, robustness, and privacy and security.

A sociotechnical system refers to how human and technical elements connect with each other \cite{cooperSociotechnicalSystems1971}. 
Computer systems including automated decision systems are not isolated and neutral tools but products of their sociotechnical context \cite{krollFallacyInscrutability2018, dignumResponsibleAutonomy2017}.
The sociotechnical context of an algorithm includes identification of stakeholders encompassing but not limited to the people that might be adversely affected by the algorithm \cite{brownAlgorithmAuditScoring2021}, the wider context that the algorithm is used in including intended uses and operators \cite{krafftActionOrientedAIPolicy2021} as well as the development phase of the system \cite{brownAlgorithmAuditScoring2021, rajiClosingAIAccountability2020} as the sociotechnical context should be considered from the very beginning of the system design \cite{krollFallacyInscrutability2018}. Human oversight, potential harms, and legal responsibility are necessarily discussed within the sociotechnical context \cite{slotaManyHandsMake2021}. Existing sociotechnical issues or not accounting for the sociotechnical context may result in biased algorithms \cite{selbstFairnessAbstractionSociotechnical2019, mehrabiSurveyBiasFairness2021}. Furthermore, ethics and regulations are attached to a systems' use in its context rather than its mere technical specifications \cite{krollFallacyInscrutability2018, brownAlgorithmAuditScoring2021, maasAligningAIRegulation2021}. The proper balance when faced with tradeoff (e.g., privacy versus statistical accuracy, explainability versus intellectual property rights) also depends on the sociotechnical context in which the systems operate \cite{GuidanceAIAuditing}.

Interpretability and explainability are often used interchangeably in the literature and their definitions, particularly the differences between the two, are not agreed upon. Many studies \cite{dosilovicExplainableArtificialIntelligence2018, gilpinExplainingExplanationsOverview2018, marcinkevicsInterpretabilityExplainabilityMachine2020, liptonMythosModelInterpretability2018, erasmusWhatInterpretability2021} recognize the lack of rigor regarding their definitions and what it means to be explainable and interpretable. Gall \cite{gallMachineLearningExplainability2018} states that interpretability is concerned with the ability to discern how the internal mechanics of the algorithm work whereas explainability is concerned with the ability to elucidate in human terms why and how these internal mechanics work the way they do. Accordingly, the next section distinguishes interpretability and explainability based on the target audience: model developers and ordinary human audience, respectively. Explainability and interpretability are indispensable characteristics of accountable systems as they make it possible or easier to identify and amend mistakes \cite{GuidanceAIAuditing}, serve as a prerequisite for meaningful human oversight \cite{GuidanceAIAuditing}, and help build trust in the system \cite{angelovExplainableArtificialIntelligence2021, longoExplainableArtificialIntelligence2020}. Various regulations already entitle users with the right to explanation \cite{selbstMeaningfulInformationRight2017, edwardsEnslavingAlgorithmRight2018, barocasHiddenAssumptionsCounterfactual2020}. There are different ways of providing explanations including individual-level versus population-level explanations, model-agnostic versus model-specific explanations \cite{kazimAIAssuranceProcesses2020, koshiyamaAlgorithmAuditingSurvey2021} and explanations that can be built inside the model or can be provided via supplemental tools \cite{GuidanceAIAuditing, mokanderEthicsBasedAuditingAutomated2021}. On the other hand, explainability and interpretability might have tradeoffs with statistical accuracy \cite{GuidanceAIAuditing, angelovExplainableArtificialIntelligence2021, zicariZInspectionProcessAssess2021} and privacy \cite{koshiyamaAlgorithmAuditingSurvey2021}. 
Although most discussions on explainability and interpretability concentrate on in-processing and post-processing stage \cite{koshiyamaAlgorithmAuditingSurvey2021}, the concept also covers the datasets (e.g., the need for data dictionaries and datasheets for datasets \cite{koshiyamaAlgorithmAuditingSurvey2021}). In relation to assurance activities, a definition of AI assurance includes ensuring that the system outcomes are trustworthy and explainable \cite{batarsehSurveyArtificialIntelligence2021}. Moreover, algorithm insurance and certifications in the future might align closely with the systems' explainability and interpretability \cite{kazimAIAssuranceProcesses2020, koshiyamaAlgorithmAuditingSurvey2021}.

Transparency is used in this paper as an umbrella term that refers to the openness associated with the decisions and actions during the system lifecycle, often manifesting itself in the form of clear, comprehensive, and useful documentation. Transparency does not necessarily result in explainable or interpretable models \cite{krollFallacyInscrutability2018} and it rather covers the explanations for the processes behind the model development \cite{selbstIntuitiveAppealExplainable2018a}. To begin with, transparency includes information about the system's existence \cite{krollFallacyInscrutability2018, brownAlgorithmAuditScoring2021} and reasons for its development and deployment \cite{kazimAIAuditingImpact2021}. It further includes the transparency about the model, its use, as well as the processes that data is collected and utilized by \cite{brownAlgorithmAuditScoring2021}. Design transparency (e.g., code and data availability), documentation of design decisions, reproducibility, operational record keeping, impact assessments are considered as requirements of a transparent system \cite{krollOutliningTraceabilityPrinciple2021}. It is important that transparency-ensuring information is available as a structured and preferably standardized documentation \cite{krollOutliningTraceabilityPrinciple2021} and covers the whole lifecycle of the software \cite{najaSemanticFrameworkSupport2021}. Ensuring transparency brings harmful behavior to light \cite{weitznerInformationAccountability2008}, may help with detecting potential biases and other issues \cite{shenEverydayAlgorithmAuditing2021, loiAccountabilityUseArtificial2021}, may improve the quality of the system \cite{loiAccountabilityUseArtificial2021}, and may enhance the trust in the system \cite{dignumResponsibleAutonomy2017}. 
Transparency requires data, processes, and results to be ready for inspection and monitoring \cite{dignumResponsibleAutonomy2017}. Such transparency can be direct via public transparency or indirect via transparency to auditors \cite{loiAccountabilityUseArtificial2021}. Regulatory bodies are also considering transparency as a principle for accountable and trustworthy AI systems \cite{felzmannTransparencyYouCan2019}. Lack of transparency may also result from the existing power dynamics and may reinforce the existing power imbalances \cite{krollFallacyInscrutability2018}.
Accordingly, transparency is a necessary condition for conducting internal or external audits \cite{rajiClosingAIAccountability2020} and for a meaningful accountability relation between the actor (the owner or operator of an ADS) and the forum (e.g., the general public, stakeholders, auditors) \cite{mossAssemblingAccountabilityAlgorithmic2021, cechAgencyForumMechanisms2021}. 
On the other hand, transparency may conflict with proprietary rights \cite{slotaManyHandsMake2021, crawfordCanAlgorithmBe2016, buhmannManagingAlgorithmicAccountability2020}. It should be noted that transparency is not an end goal in itself but serves towards achieving accountability \cite{krollOutliningTraceabilityPrinciple2021, tagiouToolSupportedFramework2019}.

Nondiscrimination is closely related to the concepts of unbiasedness and fairness. Mehrabi \textit{et al.}  \cite{mehrabiSurveyBiasFairness2021} presents 23 types of bias, six types of discrimination, and 10 different definitions of fairness. The authors describe fairness as the lack of prejudice or preference towards a person or a group based on their inherent or acquired traits. Some fairness definitions are incompatible \cite{berkFairnessCriminalJustice2021} and individual definitions cannot capture the full spectrum of notions of fairness and discrimination in legal, sociological, and philosophical contexts \cite{selbstFairnessAbstractionSociotechnical2019}. 
There are also differences between legal definitions and regulations regarding what is considered discrimination in different jurisdictions such as the EU and the US \cite{wachterWhyFairnessCannot2021}. 
In this work, discrimination and unfairness are used as loose terms referring to the legally or socially unacceptable or undesired treatment of different groups and individuals.
Although (un)fairness is usually observed in the model outcomes, to avoid discriminatory outcomes, the whole lifecycle of a system must be considered as design, development, and deployment are interlinked stages \cite{mcgregorINTERNATIONALHUMANRIGHTS2019}. It also has strong implications on training datasets \cite{kazimAIAssuranceProcesses2020} as historic data from the imperfect world and unbalanced data sets may result in bias against historically marginalized and underrepresented groups, respectively \cite{mehrabiSurveyBiasFairness2021, GuidanceAIAuditing}. As moral values of the society regarding what is unfair evolves continuously \cite{kenwardMachineMoralityMoral2021} and some potential discrimination might become known only after deployment, appropriate real-time fairness monitoring mechanisms are necessary \cite{GuidanceAIAuditing}. On the other hand, there may be tradeoffs between fairness and privacy due to the processing of demographic data to measure fairness \cite{kazimAIAuditingImpact2021} and between fairness and statistical accuracy  \cite{GuidanceAIAuditing, berkFairnessCriminalJustice2021}. 
Nevertheless, fairness is highly associated with trust in the system \cite{feuerriegelFairAI2020} and fairness of ADSs is increasingly finding a place in law and regulations \cite{hoffmannWhereFairnessFails2019,  hackerTeachingFairnessArtificial2018}. The ethical solution to fairness can be achieved through creating processes and fora that are interactive and discursive \cite{buhmannManagingAlgorithmicAccountability2020}, for which a tangible accountability framework can provide great support.

Robustness implies that ADSs must be testable during its lifecycle \cite{krollOutliningTraceabilityPrinciple2021}, be safe against adversarial attacks aimed at manipulating the system \cite{krollOutliningTraceabilityPrinciple2021, koshiyamaAlgorithmAuditingSurvey2021, wangSecurityMachineLearning2019a, sharmaCERTIFAICommonFramework2020}, and be statistically accurate \cite{koshiyamaAlgorithmAuditingSurvey2021, kazimAIAssuranceProcesses2020}.
Robustness also requires that the model performs well on the test set \cite{xuRobustnessGeneralization2012} where the test data should have similar characteristics with the real-world cases. Accordingly, lack of robustness is associated with brittleness where a system fails in unexpected ways, particularly if subjected to situations that are different from what they are trained on \cite{jenkinsNextWaveArtificial2020}. Therefore, robustness is closely linked with performance in the real world, which  is linked to the assumptions that drive the development and the potentially unanticipated facts in the real world that are observed only after deployment.
Apart from the robustness of a system, a robust accountability evaluation requires the system to be accountable to an external forum \cite{mossAssemblingAccountabilityAlgorithmic2021}, for which an objective accountability framework can be of assistance.

Privacy and security, in general, refer to the privacy and security of personal information but also include privacy of other critical data and security of the system against malicious actors \cite{GuidanceAIAuditing}. 
The discussion around privacy should focus on the appropriate use of information rather than restricting access, particularly in the context of accountability \cite{weitznerInformationAccountability2008}. Security refers to jeopardizing a system's confidentiality, integrity, or availability via intrusion \cite{strohmayerTrustAbusabilityToolkit2021}. Privacy and security are, then, also concerned with who are authorized to use the system \cite{brownAlgorithmAuditScoring2021} and should prevent personal or critical data leakage \cite{koshiyamaAlgorithmAuditingSurvey2021}. Privacy by design \cite{gursesEngineeringPrivacyDesign2011} implies consideration of privacy requirements during the whole lifecycle of the system \cite{krollFallacyInscrutability2018} including the privacy concerns for the data on which machine learning models are trained \cite{GuidanceAIAuditing}. 
Overall, appropriate documentation of the system and its components is necessary for inspecting the system for privacy and security concerns \cite{krollFallacyInscrutability2018}. On the other hand, privacy may conflict with transparency \cite{weitznerInformationAccountability2008}. However, the right to privacy is recognized and enforced by many laws and regulations \cite{custersComparisonDataProtection2018, rustadGlobalDataPrivacy2019}. To ensure privacy, in addition to adversarial testing \cite{rajiClosingAIAccountability2020} and privacy-enhancing techniques in such as differentially private machine learning \cite{gongSurveyDifferentiallyPrivate2020} and federated learning \cite{liFederatedLearningChallenges2020}, there are mitigation strategies like anonymization \cite{koshiyamaAlgorithmAuditingSurvey2021} and assurance procedures such as impact assessments related to data privacy and protection \cite{koshiyamaAlgorithmAuditingSurvey2021, biekerProcessDataProtection2016}. Finally, security, in addition to following standard security principles and procedures during development, also requires the real-time monitoring of security risks in deployment \cite{kazimAIAuditingImpact2021}.

In summary, there is a plurality of perspectives and foci regarding algorithmic accountability and auditing algorithms. However, the literature lacks an unifying accountability benchmark. Section \ref{sec:sc} section presents the system accountability benchmark and system cards. System accountability benchmark is a framework for evaluating machine learning-based decision aiding systems with respect to relevant regulations, industry standards, and societal expectations. System cards are resulting scorecards when particular ADSs are subjected to evaluation based on the proposed system accountability benchmark.

\section{Methodology}\label{sec:met}

The system accountability benchmark and the associated system cards have been developed in a two-stage process. In the first stage, an integrative literature review was conducted to identify and organize concepts and criteria in a framework. In the second stage, an expert focus study was conducted to improve the framework and additional feedback were welcomed during an asynchronous discussion period.

The integrative literature review approach is used to generate new frameworks and perspectives via reviewing, critiquing, and synthesizing the literature \cite{torracoWritingIntegrativeLiterature2016}. This paper conducted an integrative review on the emerging field of AI Accountability to create a holistic conceptualization and consequently an accountability framework for AI-based ADSs. The literature review was performed by first identifying several relevant keywords (e.g., AI Ethics, Accountable AI) and then expanding the list of keywords as identified documents are processed. The keywords were utilized for web searches made on prominent search engines, academic or otherwise (e.g., Google, Google Scholar, Scopus, Web of Science). Similarly, a snowball sampling approach was employed to identify other documents that cite or cited by the identified documents. News, technical reports, white papers, and other documents were also considered in addition to academic articles. It should be noted that a formal systematic review methodology was not necessarily followed as the goal is to systematize the knowledge within a well-representative framework rather than presenting a detailed and fully complete overview of the past literature or providing conclusive evidence on specific research hypotheses. 

As the literature was read, the common concepts and themes emerged. The literature was then processed with a critical lens informed by the emerging concepts, themes, and categorizations as well as the goal of the study that is to develop an accountability framework. As synthesis is a creative pursuit rather than an information dump \cite{torracoWritingIntegrativeLiterature2016}, the authors utilized the breadth and depth of their own insights to achieve a better understanding and organization of the topic. As a result, the system accountability benchmark, identifying a set of essential and representative accountability criteria and presenting them with an appropriate categorization, was developed.

As the authors' own interpretation of and the insight into the literature may bias the resulting artifact, the system accountability benchmark, external feedback was sought. A focus group study was conducted with a purposefully selected and demographically diverse group from academia, industry, government with expertise in law, public policy, and technology. The same group serves as the community advisory board for the broader research project, therefore, were familiar with the overall research objectives. The developed system accountability benchmark was shared with the focus group members in advance alongside a discussion on relevant background information and criteria descriptions. Their responses to the questions listed below were discussed during the focus group study and the benchmark was revised accordingly.

\begin{enumerate}
    \item Do you find the general organization of the framework appropriate? If not, how can we improve?
    
    \item Do you find the list of rows appropriate? If not, how can we improve?
    
    \item Do you find the list of columns appropriate? If not, how can we improve?
    
    \item Do you find any criteria (that we have included) to be unsuitable for inclusion in the benchmark? If yes, which ones and why?
    
    \item Do you find any criteria whose removal would not deteriorate the benchmark? If yes, which ones and why?
    
    \item Can you think of any criteria (that we have not included) that must be included in the benchmark? 

    \item Do you find any criteria to be misplaced in the framework?

    \item Do you have any other comments on the criteria?

    \item Do you find the visualization of system cards effective and efficient in communicating the results of an algorithm audit based on the benchmark? 
\end{enumerate}

The criteria presented in the framework neither quantitatively measure specific phenomena themselves nor consist of items measuring the same phenomena. Rather, they aim to collectively represent the accountability concerns in an adequate manner to assist with ethical assessments and audits of AI-based ADSs. Therefore, content validity, which is concerned with the relevance and representativeness of an instrument, is appropriate for validating the accountability benchmark \cite{lynnDeterminationQuantificationContent1986}. Creswell and Miller \cite{creswellDeterminingValidityQualitative2000} present nine different methods for validating qualitative research findings based on the validation lens (lens of the researcher, study participants, or people external to the study) and paradigm assumptions (postpositivist or systematic paradigm, constructivist paradigm, critical paradigm). This study employs the lens of study participants (i.e., the focus group/community advisory board) and a systematic paradigm, hence the recommended validation method of member checking \cite{creswellDeterminingValidityQualitative2000, lincolnNaturalisticInquiry2022}. 

To ensure and improve the content validity of the accountability benchmark with a member checking approach, an asynchronous discussion period followed the focus group. During this period, feedback from the community advisory board is continued to be sought to ensure content validity. Specifically, the system accountability benchmark was shared with the individual members of the community advisory board. For each set of criteria as defined by rows and columns of the benchmark (i.e., for each of the 16 sets resulting from four rows and four columns), each member is asked to respond to two questions. 
The first question asks whether the included criteria in each set are essential for the corresponding row and column intersection. A negative response indicates that the corresponding set contains inessential/unimportant criteria. 
The second question asks whether the included criteria in each set adequately represent all relevant and critical accountability concerns for the corresponding row and column intersection. A negative response indicates that the corresponding set lacks some other essential criteria.
Then, the authors followed up with the individual respondents to elaborate on their responses and discuss changes to satisfy their expert feedback. The benchmark was revised once again based on the described member checking procedure. Therefore, the accountability benchmark is considered to be justified for content validity.

\section{System Accountability Benchmark and System Cards}\label{sec:sc}
The system accountability benchmark consists of 56 criteria organized within a framework of four-by-four matrix. The rows of the matrix relate to the aspects of the software and the columns relate to the categories of the accountability criteria. The four aspects related to the software are \textit{Data}, \textit{Model}, \textit{Code}, and \textit{System}. The categories of the criteria are \textit{Development}, \textit{Assessment}, \textit{Mitigation}, and \textit{Assurance}.

\begin{table*}
\centering
  \caption{System Accountability Benchmark Criteria}
  \label{tab:criteria}

\setlength{\tabcolsep}{4pt}
\footnotesize
\begin{tabular}{l|lr|lr|lr|lr}
\cline{2-9}
                        & Development                   & \#                    & Assessment           & \#                    & Mitigation               & \#   & Assurance                    & \#   \\ \hline
\multirow{5}{*}{Data}   & Data Dictionary               & C111                  & Privacy, Data        & C211                  & Anonymization            & C311 & Data Protection              & C411 \\
                        & Datasheet, Collection Process & C112                  & Fairness, Data       & C212                  & Security                 & C312 & Datasheet, Maintenance       & C412 \\
                        & Datasheet, Composition        & C113                  & Quality, Labels      & C213                  &                          &      & Datasheet, Uses              & C413 \\
                        & Datasheet, Motivation         & C114                  & Inspectability       & C214                  &                          &      &                              &      \\
                        & Datasheet, Preprocessing      & C115                  &                      &                       &                          &      &                              &      \\ \hline
\multirow{5}{*}{Model}  & Reproducibility, Model        & C121                  & Interpretability     & C221                  & Adversarial, Training    & C321 & Privacy, Model               & C421 \\
                        & Design Transparency, Model    & C122                  & Fairness, Model      & C222                  & Explanations, Mitigation & C322 & Uses, Model                  & C422 \\
                        & Documentation, Model          & C123                  & Testing, Adversarial & C223                  & Fairness, Mitigation     & C323 & Documentation, Capabilities  & C423 \\
                        & Selection, Model              & C124                  &                      &                       & Privacy, Training        & C324 & Explainability               & C424 \\
                        &                               &                       &                      &                       &                          &      &                              &      \\ \hline
\multirow{5}{*}{Code}   & Reproducibility, Code         & C131                  & Privacy, Code        & C231                  & Review, Code             & C331 & Certification, Developer     & C431 \\
                        & Design Transparency, Code     & C132                  & Security, Code       & C232                  & Diversity, Team          & C332 & Due Diligence                & C432 \\
                        & Documentation, Code           & C133                  & Testing Cards        & C233                  &                          &      &                              &      \\
                        &                               &                       &                      &                       &                          &      &                              &      \\
                        &                               &                       &                      &                       &                          &      &                              &      \\ \hline
\multirow{6}{*}{System} & Documentation, Development    & C141                  & Awareness, Public    & C241                  & Monitoring, Fairness     & C341 & Record Keeping, Operational  & C441 \\
                        & Plans, Maintenance            & C142                  & Risk, Humans         & C242                  & Monitoring, Performance  & C342 & Uses, System                 & C442 \\
                        &                               &                       & Training, Operator   & C243                  & Oversight, Human         & C343 & Documentation, Acceptability & C443 \\
                        &                               &                       & Accuracy, System     & C244                  & Harms, Remedies          & C344 & Insurance                    & C444 \\
                        &                               & \multicolumn{1}{l|}{} &                      & \multicolumn{1}{l|}{} & Mechanism, Feedback      & C345 & Rating, Risk                 & C445 \\
                        &                               &                       &                      &                       & Security                 & C346 &                              &      \\ \cline{2-9} 
\end{tabular}                                 
\end{table*}

\textit{Data} refers to the aspects related to the properties and characteristics of the data that the model learns from or works with. \textit{Model} refers to the properties and behavior of the machine learning-based decision model that the system utilizes. \textit{Code} refers to the actual source code underlying the system, including the code surrounding the decision model in its development and use. Finally, \textit{System} refers to the software and its socio-technical context as a whole.

\textit{Development} covers the criteria related to the operational, and informational record keeping for the development and use of the software. \textit{Assessment} covers the criteria that involve estimating the abilities or qualities of the system and its components. \textit{Mitigation} covers the criteria that can be utilized to prevent or mitigate potential shortcomings detected in \textit{Assessment}. Finally, \textit{Assurance} covers the criteria that aim to provide guarantees regarding the software and its use.

The evaluation of the criteria can be performed by humans manually or by computers automatically. In both cases, each criterion needs to be evaluated based on objective standards and specific guidelines to be established and continuously improved by future work. The outcome of the evaluation can be binary (i.e., satisfactory or not) or on a Likert scale (e.g., poor, fair, good, very good, excellent).

\begin{figure}[h]
  \centering
  \includegraphics[width=0.865\linewidth]{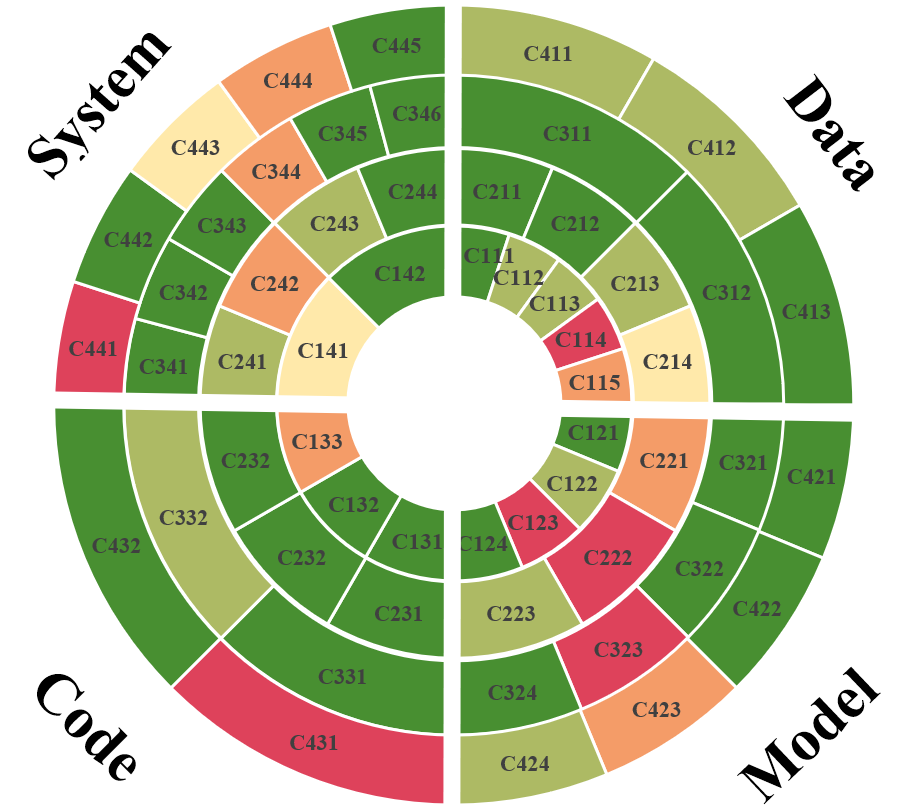}
  \caption{\textbf{Sample System Card.} Four concentric circles where each circle corresponds to a column of the framework. Each criterion is denoted by an arc within its respective circle. The color of each arc denotes the evaluation outcome for the respective criteria, from worst (shown in red color) to best (shown in green color).}
  \label{fig:viz}
\end{figure}

A system card is the overall outcome of the evaluation for a specific ADS. A system card is visualized as four concentric circles. Each circle corresponds to a column in the framework for system accountability benchmark. The circles from innermost to outermost follow the order of the columns, roughly reflective of the accountability issues in the lifecycle of an ADS. The circles are divided into four quarters with each quarter corresponding to a row of the framework. Its clockwise order follows the order of the rows, corresponding to how building blocks merge together to form an ADS, with the last quarter corresponding to the overall system. Each criterion is denoted by an arc within its respective circle and quarter. The color of each arc denotes the evaluation outcome for the respective criterion, with the worst outcome denoted in red, the best outcome denoted in green, and other outcomes in the case of a Likert scale denoted with divergent colors between red and green. Figure \ref{fig:viz} demonstrates the visualization of a system card. 

In the remainder of this section, the accountability criteria belonging to each category are discussed.

\subsection{Development}

\textit{Data Dictionary} criterion refers to the existence of data dictionaries. The data dictionary should cover training, validation, and test datasets as well as any other data that play a role in the development or use of the system. A data dictionary contains information on tables and their fields, typically including their meaning, source, relationship to other data, format, scale, and allowed values \cite{uhrowczikDataDictionaryDirectories1973, ibmIBMDictionaryComputing1993}.
The next four criteria are taken from the \textit{Datasheets} proposed for datasets by Gebru \textit{et al.} \cite{gebruDatasheetsDatasets2021}. \textit{Collection Process} aims to evoke information on how the data is collected and help other researchers collect similar data. \textit{Composition} aims to describe the composition of the dataset including aspects related to its content including relationships, recommended data splits, and sensitive information. This criterion also includes the dataset size which is shown to be a powerful signal of assumed quality \cite{waggonerBigDataTrust2019}. \textit{Motivation} aims to express the reasons for creating the dataset. \textit{Preprocessing} aims to report on if and how the dataset is processed (e.g., cleaned, transformed, labeled).

\textit{Reproducibility, Model} criterion refers to the existence of mechanisms ensuring the reproducibility of model results.
\textit{Design Transparency, Model} criterion refers to the documentation of decisions and actions related to the design and development of the model such that an issue encountered in later stages can be traced to the specific decisions and actions as well as the responsible actors. Such documentation may also assist future decision-making. 
\textit{Documentation, Model} criterion refers to the existence and adequacy of documents describing the model, akin to the model cards \cite{mitchellModelCardsModel2019}. \textit{Selection, Model} criterion refers to the existence and adequacy of documents describing the model selection process, justification of the selected model, and reasoning for how the selected model is a good fit for the particular purpose of the system.

\textit{Reproducibility, Code} criterion refers to the existence of mechanisms that allow the reproducibility of the code results.
\textit{Design Transparency, Code} criterion is similar to \textit{Design Transparency, Model}, yet it focuses on decisions regarding the code design rather than the model. 
\textit{Documentation, Code} criterion refers to the existence of a well-organized and well-documented codebase. The documentation should clearly describe each code piece, the organization of the whole codebase, and interrelations between different code pieces. Such documentation would decrease the risk of mistakes in development and increase the ease of developing improved or modified versions of the system.

\textit{Documentation, Development} criterion refers to the existence of documents that describe and tailor the whole development lifecycle of the software from ideation and problem understanding to deployment and maintenance.
\textit{Plans, Maintenance} criterion refers to the existence of detailed and actionable plans for the system's maintenance and updating, including plans for unforeseen potential changes that may become necessary or desired based on environmental and contextual changes, system's real-world performance, or stakeholder feedback.

\subsection{Assessment}

\textit{Privacy, Data} criterion is concerned with respecting the data privacy regulations and best practices and not including personally identifiable information (PII) or information such that PII can be reconstructed except for the cases where relevant regulations such as GDPR \cite{EURLex32016R0679EURLex} allow it.
\textit{Fairness, Data} criterion aims to ensure that the training data is free from bias. The input data should not contain any protected attribute (e.g., race, gender) as predictors unless it is of specific use to avoid discrimination against a protected group. The use of other variables that might serve as proxies to the protected attributes is also subject to an appropriate statistical test for input data fairness.
\textit{Quality, Labels} criterion is concerned with whether the labels for individual data instances are accurate, verified, and of good quality.
\textit{Inspectability} criterion refers to the availability of the infrastructure and tools to easily access and observe datasets that are employed or generated during the system's development and use. Such tools should integrate data dictionary capabilities, complementing the data documentation criterion.

\textit{Interpretability} criterion refers to whether the model allows developers and other technical experts to obtain explanations from it regarding how specific decisions are made.
\textit{Fairness, Model} criterion is concerned with the potential discrimination in models’ outputs against protected groups (e.g., based on race or gender).
There are multiple mutually exclusive formal criteria for fairness \cite{kleinbergInherentTradeOffsFair2016, barocasFairnessMachineLearning2019, mehrabiSurveyBiasFairness2021} and the choice of fairness metric depends on the value systems of the society and the sociotechnical context. Its evaluation is based on the model performance with respect to appropriate fairness measurement choices.
\textit{Testing, Adversarial} is concerned with providing the model with intentionally designed inputs that aim to cause the model to make a mistake. For instance, in image classification tasks, a slight change that is sometimes invisible to immediate human perception, can cause those images to be misclassified. 
Another type of adversarial attack that is worthy of noting is data poisoning \cite{wangDataPoisoningAttacks2018}. Data poisoning occurs usually in systems that employ a continuous learning approach (i.e., the models that learn from the new data as they become available, possibly including the input-output pairs obtained through the model's use in the real world). A malign actor can provide such data to a model, on purpose, to destabilize the model, worsen its performance, or manipulate it to produce certain outcomes for certain inputs. 
A model, therefore, must be tested against such adversarial attacks. Adversarial tests are employed to help evaluate this criterion.

\textit{Privacy, Code}  criterion refers to the respect for the confidentiality of user data and other sensitive data. The code should not allow for data leakage including those that can be obtained by reverse engineering (e.g., leakage of training data).  
\textit{Security, Code}  criterion refers to the overall security of the software against malicious attacks including those that aim to steal information, manipulate the software, or make the software unavailable. This criterion is closely related to information technology (IT) security. Therefore, the standards should be established with the involvement of experts in IT security.
In practice, these two criteria are interwoven and are examined together.
\textit{Testing Cards} criterion refers to whether there exist mechanisms and designs for testing the code. It should allow for testing the code’s behavior as a whole and the behavior of its individual parts. A testing card (e.g., test sheets \cite{atkinsonSoftwareTestingUsing2010}) should include information on the design of various tests (e.g., unit tests, system tests) and the results of those tests, as well as information on the code coverage. Such testing cards can assist with the evaluation of this criterion.

\textit{Awareness, Public} is concerned with ensuring an appropriate level of familiarity and knowledge about the system's existence, objectives, and mechanisms for the public in general and the decision subjects in particular. It should also publicize the system's performance in relevant accuracy and fairness metrics. Such informed awareness enables stakeholder feedback and participation regarding the system's use and potential modifications. 
\textit{Risk, Humans} criterion is concerned with the potential risks of the deployed software on the rights and freedoms of individuals. At a bare minimum, its impacts must be contemplated with potential risks to the rights defined in the Universal Declaration of Human Rights \cite{unitednationsUniversalDeclarationHuman1948} including but not limited to the right to life, liberty and personal security; rights to equality, marriage and family, education, fair public hearing; and freedom from arbitrary arrest. An ADS that violates any human right cannot satisfy this criterion. 
\textit{Training, Operator} criterion refers to whether the operators of the system 
have received adequate training on the nature and limitations of the model. Such training, possibly tailored towards different use cases, should be available alongside the software. 
\textit{Accuracy, System} criterion refers to the appropriateness of employed accuracy metrics in evaluating a systems’ performance, whether the evaluation is performed on a suitable test set, as well as the acceptability of the accuracy level.

\subsection{Mitigation}

\textit{Anonymization} is a mitigation strategy for removing personally identifiable information or using other techniques such as aggregation to ensure compliance with privacy regulations. It is evaluated based on the need for and the existence of anonymization mechanisms. \textit{Security} criterion refers to the secure communication techniques regarding how the data is securely transmitted and/or stored, in general and before anonymization/aggregation.

\textit{Adversarial, Training} is a mitigation technique, especially for when \textit{Testing, Adversarial} criterion is not satisfied. With adversarial training \cite{wangSecurityMachineLearning2019a}, the model is trained on adversarial samples to improve robustness against adversarial attacks.
\textit{Explanations, Mitigation} criterion refers to the utilization of supplementary explainability tools and techniques (e.g., surrogate explanations \cite{ribeiroWhyShouldTrust2016a, lundbergUnifiedApproachInterpreting2017}) in case the original model does not have inherent mechanisms for providing explanations.
\textit{Fairness, Mitigation} criterion refers to the utilization of techniques in pre-processing, in-processing, or post-processing stages to ensure fairness in the model outcomes \cite{mehrabiSurveyBiasFairness2021}.
\textit{Privacy, Training} criterion refers to the utilization of privacy-preserving machine learning techniques such as federated learning \cite{liFederatedLearningChallenges2020} and differential privacy \cite{gongSurveyDifferentiallyPrivate2020}. These four criteria are evaluated based on the need for and the existence of respective mitigation mechanisms.

\textit{Review, Code} criterion refers to the employment of code review practices during the development. This criterion is evaluated based on the existence, quality, and application of well-established code review practices that include code reviewers other than the original authors of the code. \textit{Diversity, Team} criterion refers to the diversity of the developer team. This work employs social category diversity based on the representation of diverse demographic groups \cite{liangEffectTeamDiversity2007}.

\textit{Monitoring, Fairness} and \textit{Monitoring, Performance} refer to the existence of infrastructure and mechanisms to monitor the fairness metrics and the accuracy of the machine learning model, respectively, in its real-world use.
\textit{Oversight, Human} criterion refers to the principle that the existence of automated systems should not reduce the power and responsibility of humans either in explicit or implicit ways \cite{mcgregorINTERNATIONALHUMANRIGHTS2019}. There are different levels of human involvement (i.e., human-in-the-loop where human consent is necessary by default, human-on-the-loop where humans can overturn the outcome of a decision-aiding system, and human-out-of-the-loop where no human oversight exists \cite{wieringaWhatAccountWhen2020}). A human-out-of-the-loop system cannot satisfy this criterion for any type of application. For higher-risk applications, humans should oversee the algorithmic decisions and modify them as appropriate and necessary. As the risk increases, the necessary human involvement should increase. However, even low-risk applications should have a mechanism where humans can overrule the decision made by the system in line with appropriate policy guidelines. This criterion is evaluated based on the existence and appropriate design of the policies and necessary mechanisms for human involvement.
\textit{Harms, Remedies} criterion aims to ensure that, even in the absence of any negligence or bad faith, appropriate remedies are provided for unintended or unexpected harms. If and when such harms occur, appropriate policies and mechanisms should be readily available to establish the nature of remedies with respect to the seriousness of the harm. This criterion is evaluated based on the availability, sufficiency, and effectiveness of the remedies.
\textit{Mechanism, Feedback} criterion refers to whether appropriate and effective mechanisms are in place for enabling and receiving feedback from different stakeholders including the decision subjects. It is also concerned with the readiness and ability to constructively respond to such feedback, incorporating any potential changes to the system swiftly, and communicating any necessary information to the system's developers, operators, and relevant stakeholders.
\textit{Security} criterion is concerned with whether system security tools and techniques are in place and effective agains potential security threats. 

\subsection{Assurance}

\textit{Data Protection} criterion is concerned with whether data protection impact assessments \cite{DataProtectionImpact2021, biekerProcessDataProtection2016} are prepared for the system. This criterion is evaluated based on the availability and adequacy of impact assessments with respect to relevant regulations and best practices.
\textit{Datasheets, Maintenance} criterion aims to ensure dataset maintenance plans are in place and the plan allows for future communication regarding the dataset. 
\textit{Datasheets, Uses} aims to clarify tasks for which the dataset can be used as well as the tasks that the dataset should not be used for.

\textit{Privacy, Model} criterion refers to whether appropriate privacy precautions are in place with respect to the model architecture and behavior. \textit{Uses, Model} criterion refers to the alignment of the model's actual use in the real world to the purpose that it is originally intended for. 
\textit{Documentation, Capabilities} criterion refers to the existence and clarity of the documents that provide detailed information on the capabilities of the model (e.g., tasks that it can successfully perform, conditions that are necessary for it to perform, intended users). It is also important that the documentation explicitly describes the tasks and situations for which the model is not suitable to use (i.e., its limitations). 
\textit{Explainability} criterion refers to the ease of understanding and interpretation for the ordinary human audience regarding how the model arrives at its decisions. A model should be self-explanatory or should provide additional tools for explaining how it produces its outputs, both for the general model behavior and for individual cases. Explainability cannot be satisfied merely by explicating the calculations but requires the audience to be able to make counterfactual claims about the model's behavior \cite{wangCounterfactualExplanationsExplainable2021}.

\textit{Certification, Developer} criterion refers to whether developers of the system (i.e., engineers, designers, testers). have appropriate and necessary certificates. 
\textit{Due Diligence} refer to whether licensing issues, open source clearance, and other ownership information are clearly defined, agreed-upon, and documented.

\textit{Record Keeping, Operational} criterion refers to the existence of technological infrastructures that provides logging capabilities for the duration of the system’s development and use. The records must clearly indicate inputs, outputs, model files as well as information on the use of the system such as timestamps and operators.
\textit{Uses, System} criterion refers to the alignment of the system's actual use in the real world to the purpose that the system is originally intended for.
\textit{Documentation, Acceptability} criterion refers to the existence and completeness of the documents that elaborate and certify the acceptance criteria for the system and the conditions necessary to satisfy those criteria. A set of fixed acceptance criteria is neither established in the literature nor proposed in this work. However, the acceptance criteria should be unambiguous, consistent, and comprehensive (i.e., cover relevant aspects related to functionality, performance, interface, security, and software safety \cite{wallaceGuideSoftwareAcceptance1990}). The acceptance criteria should be developed with the involvement of key stakeholders and must account for the project requirements \cite{modelingandsimulationenterpriseVerificationValidationAccreditation2011}.
\textit{Insurance} criterion is concerned with whether the system is insured for liability. This criterion is evaluated based on the existence and coverage of the insurance.
\textit{Rating, Risk} criterion is concerned with risk ratings assigned to the system by independent risk agencies, as some regulations (e.g., GDPR \cite{EURLex32016R0679EURLex} and Artificial Intelligence Act \cite{EURLex52021PC0206EURLex}) have a tiered system of expectations based on the risk involved with the technology. The main input for this criterion is sourced from risk agencies.

\section{Discussion}\label{sec:dis}

This section discusses the concept of accountability in the context of AI-assisted decision-making, the value proposition for the system accountability benchmark, the applicability of the benchmark, and measurement of the accountability criteria.

The name and the scope of the emerging field that focuses on sociotechnical context, interpretability and explainability, transparency, nondiscrimination, robustness, and privacy and security in the context of AI have not been well defined or agreed upon. The field is generally referred to as AI Ethics, Trustworthy AI, Responsible AI, or AI Accountability. This paper employs the term \textit{accountability} to highlight AI systems' broader sociotechnical context and approach them as tools shaped and controlled by humans. This choice stems from a wider description of accountability rather than its narrower definitions focusing on accepting responsibility. With respect to this, when referring to accountability, this paper generally follows Boven's accountability framework \cite{bovensAnalysingAssessingAccountability2007} which requires a relationship between an actor (e.g., an ADS and its developers/operators) that is obliged to elucidate and justify its conduct (e.g., via system cards) and an authoritative forum that is able to make inquiries and render binding decisions (e.g., via algorithm audits and legal/social empowerment). 

The value of the proposed accountability benchmark lies in its exhaustiveness in terms of criteria covering the components, aspects, and lifecycle of an ADS; in its ability to guide the development, use, and audits of an ADS; and the motivation it provides for further enhancing itself and algorithm audits. The benchmark not only covers data and  machine learning models but also covers the surrounding code and system as a whole since a system's characteristics may be different than an amalgamation of the characteristics of its parts. In addition, a comprehensive spectrum of an ADS's relations and interactions with developers, operators, and stakeholders are covered. Moreover, the included criteria goes beyond current issues and foresee future developments in the field such as insurance and risk ratings for AI systems. Stemming from its comprehensiveness and real-world orientation, in addition to providing a checklist for auditors, the benchmark can inform AI developers to be precautious and take necessary actions to ensure accountability. As a flexible framework in terms of how individual criteria can be assessed and being open to well-informed changes to the list of included criteria, the benchmark further encourages developing its newer, expanded, or specialized versions.

While many of the individual criteria may have more general applicability, the benchmark as a whole is typically applicable to ADSs that are created within a supervised machine learning paradigm. Many criteria related to data imply a machine learning paradigm and the criteria related to the training procedures imply a supervised learning paradigm. Therefore, the benchmark is not very suitable for unsupervised learning tasks such as clustering or rule-based systems that are not powered by underlying data. Moreover, the need for using a comprehensive benchmark such as the system accountability benchmark to audit all kinds of machine learning-supported ADSs is subject to debate. Some decisions (e.g., those involving criminal justice or healthcare) are more critical than others (e.g., new product recommendations). Therefore, not all ADSs call for an extensive audit based on this benchmark or at least do not require the same stiffness when evaluating ADSs for each criterion.

The questions of how each criterion should be assessed (i.e., measured) have no easy answers. First, for some criteria, there are multiple ways to assess. Furthermore, for instance, in the case of assessing fairness, different fairness notions are not necessarily compatible and choosing a fairness notion depends on the value system rather than an objective comparison. Consequently, choosing the most appropriate assessment method may depend on the characteristics of the decision task. Second, for some criteria, an appropriate assessment method may not be readily available since such criteria are more forward-looking and not yet mature, as is the case with AI insurance or team diversity. Consequently, more collective academic and practical knowledge need to be accumulated before devising measurement methods for such criteria. Third, some criteria can be evaluated only qualitatively as the artifacts related to those criteria cannot be reasonably reduced to numbers, as is the case with most criteria involving documentations. Consequently, the evaluation of those criteria are highly coupled to the auditors and can be addressed in relation to the design and elaboration of auditing processes. Finally, it is not feasible to reduce the binary or Likert-scale outcomes for individual criteria to a single score or a pass/fail decision for the whole system without considering the social and legal context of a specific ADS.

\section{Conclusion}\label{sec:co}

The contribution of this work is the proposed system accountability benchmark that produces system cards for machine learning-based automated decision systems in various functions and tasks. The entries of a system card are to be filled by relevant internal or external auditors based on a proper evaluation of the overall system and its components. Apart from serving the auditing purposes, the framework is intended to be considered during the whole lifecycle of the system to assist its design, development, and deployment in accordance with accountability principles.

Considering the evolving and cumulative nature of science in general and the infancy of the algorithmic accountability field in particular, the proposed framework for system accountability benchmark is not suggested to be the ultimate solution for all and at once. However, being a tangible and relatively specific framework, the system accountability benchmark and system cards contribute to moving the current efforts forward towards real-world impact, ignite further discussions on the topic, and motivate holding automated decision systems accountable in practice.

The system accountability benchmark paves the way for three major lines of future work that are also sequentially linked. First, further research is required to establish specific and full-fledged guidelines or automated tools to evaluate the proposed criteria.
The second line of work is towards developing mature auditing procedures that elaborate on what to audit, when to audit, who audits, and how audits should be conducted.
Finally, in light of the auditing procedure, the proposed framework alongside the specific guidelines and automated tools can be applied to generate system cards for different AI-based automated decision systems in the real world.

\section*{Acknowledgment}

The authors would like to thank Drs. Ryan Kennedy, Andrew C. Michaels, and Lydia B. Tiede for their insightful comments on an earlier draft of this manuscript.

The authors would also like to thank Jesse Bounds, Katherine A. Franco, Sharon Israel, Doug O'Brien, Maurice Singleton and other (anonymous) members of the Community Advisory Board for their feedback and contributions on developing and validating the system accountability benchmark and system cards.

This material is based upon work supported by the National Science Foundation under Award CCF-2131504. Any opinions, findings, and conclusions or recommendations expressed in this material are those of the authors and do not necessarily reflect the views of the National Science Foundation.

\bibliographystyle{IEEEtran}
\bibliography{references}

\end{document}